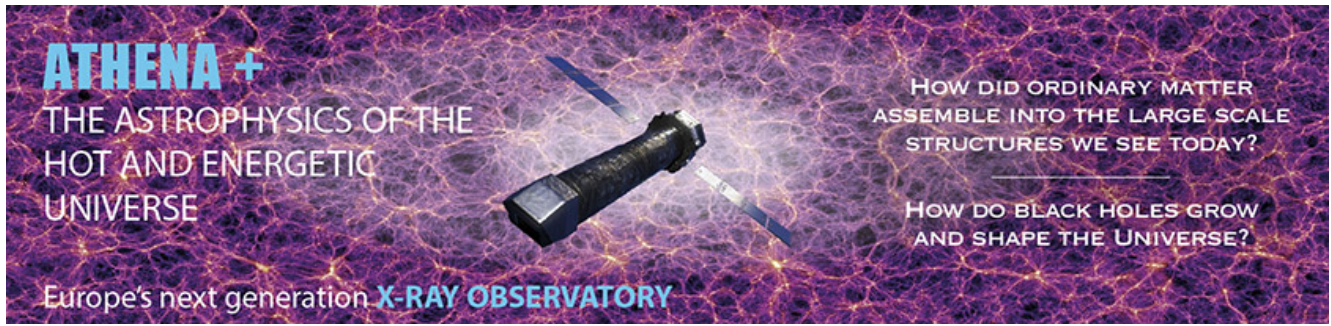

# The Hot and Energetic Universe

An *Athena+* supporting paper

# Solar system and exoplanets

Authors and contributors

## The X-ray view of the solar system

**G. Branduardi-Raymont,** K. Dennerl, M. Holmstrom, D. Koutroumpa, A. Read, A. Bhardwaj, Y. Ezoe, R. Gladstone

## X-rays and exoplanets

**S. Sciortino**, A. Maggio, G. Micela, M. Güdel, I. Pillitteri, J. Sanz-Forcada

# 1. THE X-RAY VIEW OF THE SOLAR SYSTEM: SUMMARY

*Athena+* investigations of the solar system, by giving us ever deeper insights in the complex workings of planetary magnetospheres and exospheres, will answer many of the questions left open following the pioneering work done with Chandra and XMM-Newton, and will add enormously to our understanding of the interactions of space plasmas and magnetic fields. In the case of Jupiter, *Athena+* X-IFU will unequivocally determine the species, and thus the origin (solar wind or Io's volcanoes), of the ions responsible for the soft X-ray aurora, and will test theories of ion acceleration in the planet's magnetosphere through line broadening velocity measurements. Simultaneous higher energy spectra will characterize the energy distribution of the electron population in the Jovian neighbourhood. High sensitivity observations of X-ray fluorescence from the Galilean moons will allow surface composition measurements, and studies of the Io Plasma Torus will shed light on the yet unknown mechanisms energising its X-ray emission. The X-IFU will search for X-ray aurorae on Saturn to much greater depth than it has been possible so far, and will put very sensitive constraints on Uranus' and Neptune's X-ray emissions. The non-dispersive high spectral resolution of the X-IFU will allow us to map spectroscopically Mars' extended exosphere under differing solar wind conditions and over the seasons, as well as the very extended comae of comets transiting in the Sun's neighbourhood. Solar system research with *Athena+* will also provide necessary insights to understanding the details of the charge exchange process, and to applying them to the wider context of *Athena+* main science. Solar system studies are directly applicable to a variety of astrophysical scenarios, including the investigation of exoplanet atmospheres (see section 2), and will add to the growing importance of cross-disciplinarity among astrophysics, planetary and space plasma research. Short of making X-ray observations in situ at planets and comets, *Athena+* is the only forthcoming opportunity to make significant progress in this field. Including solar system bodies as targets for *Athena+* truly adds a new dimension to the mission's science.

## 1.1. Introduction: What we know

X-ray studies of our solar system have reached a degree of maturity in the last decade, thanks to XMM-Newton and Chandra. Planets, their moons and comets have been detected and in some cases studied in detail; the significance of the Charge eXchange (CX) process has been realised (e.g. Dennerl 2010), with Solar Wind CX (SWCX) being responsible for the soft X-ray emission in most cases.

Jupiter's auroral spectrum below 2 keV is dominated by CX line emission from the encounters of highly stripped, energetic ions and $H_2$ molecules of the planet's upper atmosphere (Branduardi-Raymont et al. 2007). The origin of the ions, from the solar wind or the inner magnetosphere, i.e. Io's volcanoes, has been matter of debate, and the latter hypothesis is currently favoured. At higher energies, the X-ray spectrum is featureless, and attributed to electron bremsstrahlung. The electron component varied significantly over a 3.5 day XMM-Newton observation, probably in response to solar activity.

Scattering of solar X-rays in Jupiter's atmosphere results in further emission, free of auroral contamination at low latitudes, with a spectrum closely resembling that of the Sun, with strong iron and magnesium emission lines. Fig. 1 shows the different morphology observed in narrow energy bands: the aurorae are very evident in the band centred on the CX OVII line (top left panel) and at higher energies, where bremsstrahlung dominates (bottom panels), while a round and uniform disk is observed if we select iron lines (top right panel), characteristic of the solar coronal spectrum.

For the rest of the Jovian system, X-ray detections of Io, Europa and possibly Ganymede have been obtained with Chandra, as well as a measurement of the X-ray spectrum of the Io Plasma Torus, characterised by a puzzling very soft continuum and a line at ~ 0.57 keV, which could be from OVII with contributions from other oxygen charge state transitions.

Unlike Jupiter, Saturn has shown no evidence (yet) for auroral X-ray emission; a combination of scattering and fluorescence of solar X-rays is thought to be responsible for its disk, polar cap and ring emissions (Bhardwaj et al. 2005a,b, Branduardi-Raymont et al. 2010). Fluorescent scattering appears to take place in the upper atmospheres of both Venus (Dennerl et al. 2002) and Mars (Dennerl et al. 2006); Mars has also been found to possess an X-ray emitting exosphere. Surprisingly, the emission extends out to several planet radii; the details of this are still unclear, but the emission is probably related to Mars' atmospheric loss: its spectrum displays a remarkable richness of SWCX emission lines (Fig. 2). The exosphere of Venus is more compact, compared to Mars, due to its higher mass, which makes separating spatially the SWCX and scattered solar X-rays very challenging and feasible only during solar minimum (Dennerl 2008).

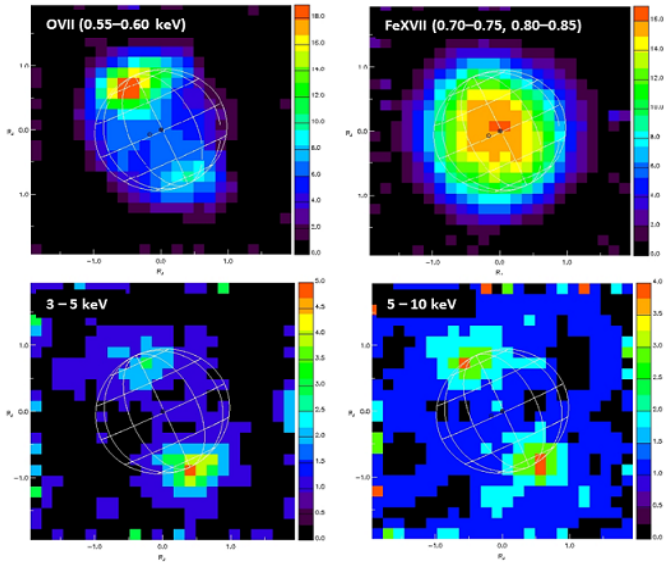
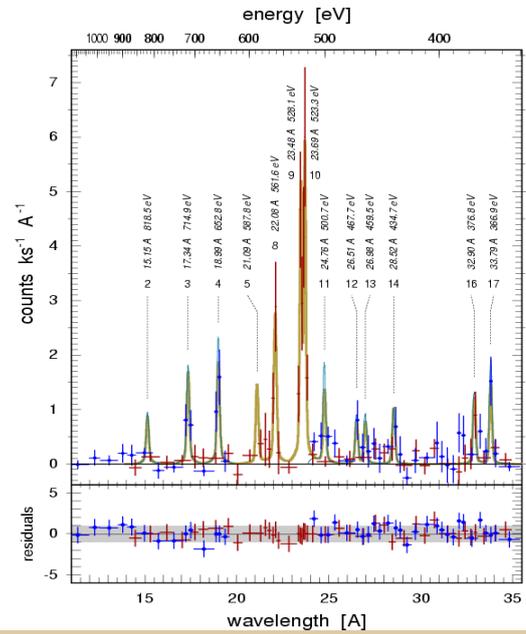

Fig. 1 – XMM-Newton images of Jupiter in narrow energy bands showing the auroral (top left and bottom panels) and disk contributions (top right) (from Branduardi-Raymont et al. 2007).

Fig. 2 – XMM-Newton RGS spectrum and model fit of Mars' disk and its halo (out to ±50" in the cross-dispersion direction from the planet's centre - from Dennerl et al. 2006).

A similar SWCX origin is attributed to X-rays from comets and the heliosphere. Comets, with their extended neutral comae, are ideal targets for SWCX: they are spectacular X-ray sources, with extremely line-rich X-ray spectra, and excellent probes of the conditions of the solar wind in the Sun's proximity (Dennerl et al. 2003, Bodewits et al. 2007). The diffuse soft X-ray CX emission of the heliosphere is recognised on one side as a signal of interest, and on the other as an unfortunate contamination in studies of the hot, diffuse plasmas in the Milky Way and beyond. SWCX has also been found to occur in the vicinity of the Earth's magnetosphere (e.g. Snowden et al. 2004) and to peak in the sub-solar magnetosheath, a region where both solar wind and neutral exospheric densities are high (Carter et al. 2011). It is now clear that CX as a soft X-ray line emission mechanism is ubiquitous throughout the Universe, and its role extends from the solar system to interstellar clouds, galactic winds and galaxy clusters (Lallement 2004).

### 1.2. What *Athena+* can do

Turning to the impact that *Athena+* will have on solar system science, in Jupiter's case high sensitivity, high resolution spectroscopy with the X-IFU will unequivocally determine the ions species (C or S, from the solar wind or Io's volcanoes) responsible for the CX auroral emission, settling the long standing issue of its origin and establishing the relative contributions, and will test theories of ion transport and acceleration in Jupiter's magnetosphere by line broadening velocity measurements.

Simultaneous spectra of Jupiter's auroral bremsstrahlung at higher energies will characterise the electrons energy distribution; observation of variability will show how the ionic and electron populations respond to changing conditions that may be related to solar wind variability (inferred from propagation of the conditions at 1 AU, or, timescales permitting, from in situ measurements by JUICE). High sensitivity and spectral resolution will be invaluable in establishing the surface composition of the Galilean satellites by X-ray fluorescence, as well as the mechanisms energising the X-ray emission from the Io Plasma Torus, through spectral and variability characterisation, also under changing solar wind conditions.

At the *Athena+* angular resolution (< 5" HEW) Jupiter's auroral emissions will be far better resolved than by XMM-Newton (~16" HEW, Fig. 1). Unlike the XMM-Newton RGS spectrum (Fig. 3), where the spectral and spatial information are combined for a moderately extended source like a planet, the non-dispersive character of X-IFU spectroscopy will enable auroral and scattered solar emissions to be individually mapped spatially and spectrally at high resolution, with the advantage that the X-IFU will offer *two orders of magnitude improved effective area* and extend the sensitivity to the softer band where the C / S ambiguity can be resolved (compare Fig. 3 for RGS in 210 ks and Fig. 4 for X-IFU, 20 ks, using the same RGS model).

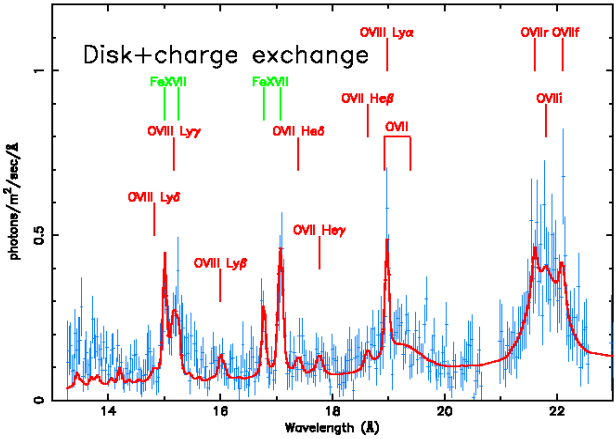 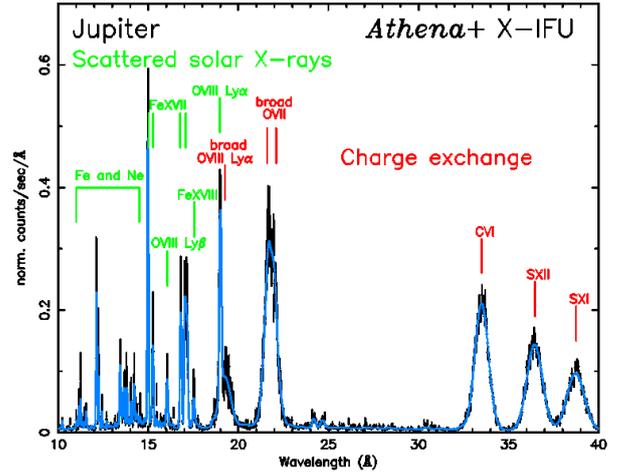

**Fig. 3** – XMM-Newton RGS spectrum of Jupiter for an exposure time of 210 ks (from Branduardi-Raymont et al. 2007).

**Fig. 4** – Simulated *Athena+* X-IFU spectrum for the same model of Fig. 3 and an exposure time of 20 ks: the wavelength coverage extends to the 30 – 40 A band allowing to resolve the C / S ambiguity. Particle background (the only relevant) is negligible and is not included.

*Athena+* X-IFU spectroscopy will push the search for auroral CX X-ray emission on Saturn to fainter limits, and attempt to establish what differentiates this giant, fast rotating planet, with its complex magnetosphere dominated by the interaction with Enceladus (which may also be a source of X-rays), from the conditions, similar in principle, on Jupiter. Observations of Uranus and Neptune with *Athena+* could provide the very first constraints on the level of their X-ray emission, and lead to detections if their emission is stronger than expected by a simple scaling from Jupiter and Saturn.

The line-rich X-ray emission from Mars' extended exosphere still has to be explored in detail and it will be of particular interest to study it under differing solar wind conditions. Benefitting from the much faster *Athena+* observations than those obtainable with the XMM-Newton RGS, exospheric and atmospheric emissions can be very effectively mapped spectroscopically by the X-IFU. Understanding the spectra obtained will be helped by the fact that the CX process leads to different line ratios than thermal plasma. This will give a global picture of Mars' extended exosphere, and reveal how it varies with season and solar wind conditions. A similar motivation would apply to Venus (although its exosphere is much smaller than that of Mars), but unfortunately this planet is too close to the Sun to be observable from L2. Any potential SWCX X-ray contribution from the Earth's magnetosphere, which however is brightest on the day-side, might be identified through line variability in the diffuse soft X-ray background, or through spectroscopic discrimination via line intensity ratios, as for the heliosphere.

Planets and comets are moving targets, so the observing strategy will have to take this into account. Comets (one or two of which are passing at ~ 1 AU on average every year) usually show very extended (several arcmin) X-ray emission. With slitless grating spectrometers, such as those on XMM-Newton and Chandra, the mixing of spectral and spatial information along the dispersion direction poses a severe challenge; in this respect, the ability of the *Athena+* X-IFU to provide non-dispersive high spectral resolution will revolutionise the observing: the X-IFU will make it possible to investigate in a straightforward way spectral changes across the X-ray emitting region (which are expected due to the superposition of emission from ions exhibiting different cross sections). X-IFU spectra will provide important data on fundamental atomic physics and on the details of the interaction between a hot plasma and a cold gas cloud (far from thermal equilibrium), information which is very difficult to gather otherwise. Sun avoidance constraints at L2 imply that comets will be observable on their way towards and away from the Sun, with a more compact X-ray appearance which will suit well the X-IFU FOV, while its high spectral resolution will be a splendid match to their line-rich spectra. For comets that may not fit in full in the X-IFU FOV, their proper motion will provide a smooth sampling of the coma, time-tagged for a direct transformation from the celestial reference frame to the comet's frame. This, complemented by the wider FOV, medium resolution spectral images provided by the WFI (e.g. providing global mapping of the bow shock) will allow us to fully exploit the potential offered by cometary SWCX X-ray emission as a probe of the solar wind, its ionic composition and speed, at varying distances from the Sun.

## 2. X-RAYS AND EXOPLANETS: SUMMARY

In the nearby known planetary systems hosting hot Jupiters, with the large effective area and the exquisite spectral resolution of *Athena+* we can search for ingress/eclipse/egress effects during planetary orbits. In a wider sample of planetary systems we can confirm/improve the statistical evidence of Star-Planet Interactions (SPI) and search for those variability features that are imprints of such interactions. *Athena+* can possibly discover unexpected spectral signatures (and their orbital modulation) of planetary atmospheres due to the host stars high energy radiation and particle emission. *Athena+* will drastically improve the knowledge of the X-ray incident radiation on exoplanets, a crucial element in order to understand the effects of atmospheric mass loss and, more generally, of the chemical and physical evolution of planet atmospheres, particularly relevant in the case of young systems. These are exciting unique contributions to exoplanetary astrophysics.

### 2.1. On-going research

Over the last decade research in extra-solar planets, or exoplanets, has been rapidly growing. Most of the results obtained so far have been based on accurate time series in the optical bandpass and accurate radial velocity measurements. Thanks to a census of about 1000 planetary systems we have discovered the, unpredicted, existence of many planets of Jupiter mass and radius in extremely close orbit (0.03-0.1 AU) as well as planetary systems with a structure vastly different from that of our solar system. Those findings have led to a deep and, still on-going, revision of the mechanism of formation and early evolution of planetary systems. Growing evidence is accumulating on the key role of high energy radiation in this process: in the very early phase at an age of 1-10 Myr star formation should be deeply affected by the intense ($10^2$-$10^3$ times the present day solar coronal emission) X-ray emission, inducing the Magneto-Rotational Instability in the (weakly) ionized circumstellar disks (e.g. Glassgold et al. 1997, Kirilov & Stefani 2013).

Furthermore, the intense stellar X-ray and UV emission up to 100 Myr will dissociate and ionize molecules in planetary thermospheres and exospheres so that light atoms escape into the interplanetary medium (cf. Güdel 2007, Penz et al. 2008). The solar wind, coronal mass ejections and flare particles may erode even the entire early planet atmosphere if no planetary magnetic field is present (Sanz-Forcada et al. 2010). Intense EUV and X-ray emission can lead to hydrodynamic escape of the atmospheres in extra-solar 'hot Jupiters'; such processes were probably important also on Venus, Earth and Mars during the first $10^9$ yr (Penz et al. 2008). The physical and chemical evolution of planetary atmospheres is a fundamental question for a full understanding of extra-solar planets.

In order to advance our knowledge in this field we need to study in detail properly selected interesting systems : they will be easily chosen from the thousands of extra-solar planets that will be known by 2028 thanks to the CoRoT, Kepler, CHEOPS and TESS space missions and other planetary search programmes. In particular, GAIA can find the young, highly eccentric systems with which we can test the predicted/expected UV and X-ray irradiation effects as a function of star-planet separation at various orbital phases.

As of today it has been possible to start studying the atmospheres of only a handful of Jupiter-like planets. Infrared observations of HD 189733b transits have shown that the planetary atmosphere contains $H_2O$, $CH_4$ and CO (Tinetti et al 2007; Swain et al. 2008); for another planet, HD 209458b, indications of silicate clouds have been found (Richardson et al. 2007). Planetary atmosphere studies based on high resolution IR spectra are among the key research topics planned for the E-ELT. Future IR studies (EChO, JWST) will strongly advance the field in the next decade and X-ray observations with *Athena+* will provide additional insights in the environment and conditions for extra-solar planets in several respects.

### 2.2. The potential of *Athena+* for exoplanet research

*Athena+* can measure both the quiescent and flare activity of a properly selected sample of stars which will be known to have planets in their habitable zones. Combined with stellar activity evolutionary trends and planetary atmospheric modeling, *Athena+* findings should conclusively show the effects of the X-ray emission of the host star on planets and provide unique insights into the atmospheric history of the potentially habitable rocky planets. Very young and distant planetary systems in their formation process (of which the first few are being found) will be of special interest: *Athena+* observations combined with advanced modeling will address the fundamental question of whether initial atmospheres can be eroded and whether such planets will evolve to dry bodies, or keep some of their initial water content.

For the physics of the hot Jupiters, their atmospheres and their evolution, *Athena+* repeated observations of nearby known systems hosting transiting planets will offer the possibility of exploring the transits in the X-ray band, as already done in the IR, optical and, recently, in the UV (Linsky et al. 2010); moreover, since hot Jupiters are orbiting very close to their host stars, *Athena+* will allow investigations of the high-energy absorption radius of the planet (cf. Fig. 5) in an atmosphere which is very likely to be expanded under the effects of the intese radiation field. In this respect, the large collecting area of *Athena+* will allow us to measure, in an unprecedented way, small variations of flux (either broad-band or in a narrow spectral range) of normal stellar systems (cf. Fig. 6). The study of photometric and, possibly, spectroscopic properties of such X-ray transits can provide constraints on models of the atmospheric evaporation, the exosphere composition and, eventually, the planetary evolution. As of today the best target for this is the hot Jupiter orbiting around HD 187933, where, by averaging over about 7 transits, the resulting light curve will allow detecting (> 3 sigma) the X-ray transit, with an amplitude down to 2-4%, in each of 4-5 time intervals spanning the full transit (see Fig. 6).

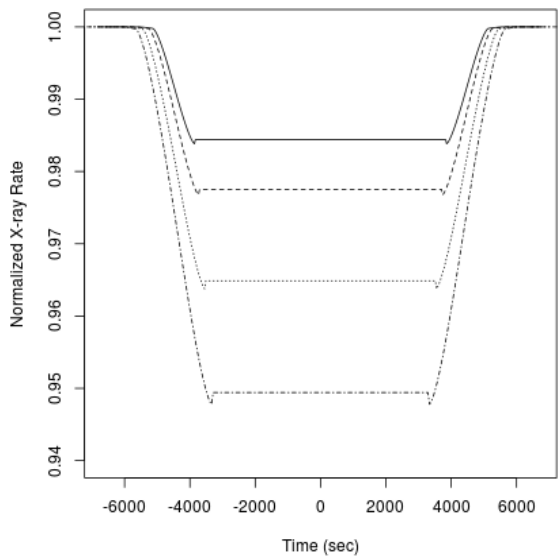 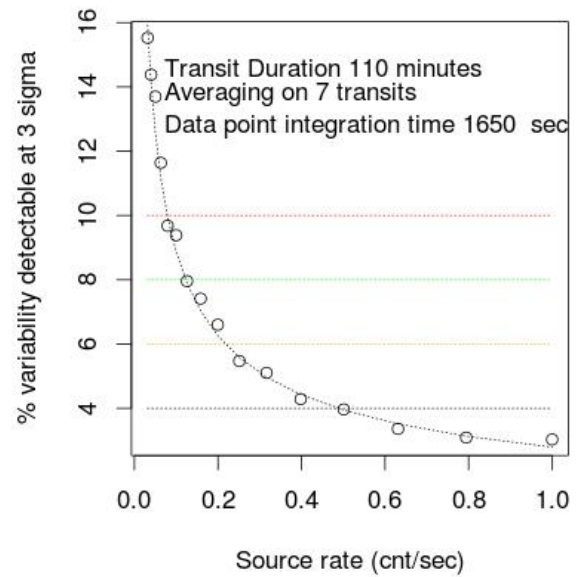

**Fig. 5:** A simple 'noise-less' model of an X-ray transit for a hot Jupiter like HD 189733a. The curves are computed for a planet X-ray obscuring radius 1, 1.2, 1.5, 1.8 (from top to bottom curves) times the optically derived radius and assume a corona described as a central disk and an outer brighter ring with a flux ratio of 1.5; the model does not include statistical or astrophysical noise. Note that a larger X-ray obscuring planet radius produces a greater intensity dimming The optically thin stellar corona produces a 'limb brightening', i.e. a ring of enhanced coronal emission. Available solar and stellar data show that the typical width of the coronal bright limb is about 10% of the stellar radius. This limb brightening produces a 'W' shape signature in the light curve (cf. Schlawin et al. 2010) which is more pronounced with higher level coronal 'limb brightening'.

**Fig. 6:** Sensitivity for an X-ray transit as a function of source intensity for a simulated transit with the values of transit time, Rplanet/Rstar and Rstar of HD 189733 (Log Lx = 28.16, d = 19 pc, orbital radius = 0.031 AU, transit duration 110 min) and an integration time of 1650 sec. The dotted curve shows the sensitivity accounting just for a very conservative X-ray background level, while the simulated sensitivity points (open circles) account also for the astrophysical noise due to the intrinsic variability of the host star's coronal emission. These results show that, in the case of an intense source (rate > 0.5 cnt/sec), variations smaller than the 4% level can be revealed at better than 3 sigma. .

The photometric variability sensitivity illustrated in Fig. 6 takes into account the known host star coronal variability acting as astrophysical noise that limits the transit detectability. For a close-in orbiting planet, like HD 187933a, we expect the atmosphere to be inflated, very likely resulting in a substantial increase of the planet's obscuring disk radius in the X-ray bandpass with respect to the optical/IR. Such an increase, as shown in Fig.5 (left), improves the detectability of the X-ray transit since, as a result, the nominal transit depth is enhanced and the duration lengthened. If the planet obscuring radius in X-rays is significantly (1.5-2 times) larger than the optical one, this can be recognized. Moreover, the data collected during the probed transits will provide very valuable repeated shadowgraphs of emitting coronal structures and can help constraining the size and temporal evolution of those structures. This is particularly relevant in the case of young host stars whose X-ray emission is known to be 2-3 orders of magnitude higher than the Sun's and whose magnetic field topology and related coronal structures can be different from the solar ones. A relevant question regards the number of stellar systems for which such a study will be feasible. Part of the answer comes from : a)

The consideration that all the ~ 106,000 RASS (ROSAT All Sky Survey) sources (at a limiting $f_X$ of ~ $2 \cdot 10^{-13}$ erg cm$^{-2}$ sec$^{-1}$) will have an *Athena+* rate > 1 cnt sec$^{-1}$, b) 40-50% of them are normal stars of spectral type F-M, c) About 14,000 RASS sources are stars listed in the Tycho catalogue, with $V_T$ < 11.5. Hence, from the X-ray perspective, there are plenty of stellar systems for which exoplanetary studies will be possible, the actual sample being composed by those systems in which planets and/or transiting planets will be discovered in the next decade.

We have also accumulated tantalizing evidence of magnetic interplay between stars and planets. By analogy with processes in close binary stars which lead to enhanced activity levels, it is thought that also giant planets might influence the stellar activity by tidal or magnetic interaction (Cuntz 2000, Saar & Cuntz 2001), although on a level weaker than in binary stars. Some indications for such Star - Planet Interactions (SPI) exist from chromospheric measurements in stars with hot Jupiters (Shkolnik et al. 2003, Fares et al. 2012), while in the X-ray bandpass such evidence is still controversial. Based on a survey of X-ray emission from stars with giant planets, Kashyap et al. (2008) have statistically shown that stars with close-in giant planets are on average up to 4 times more X-ray active than those with more distant planets. This result points to entanglements of the stellar and planetary magnetospheres. Similar results have been claimed by Scharf (2010) in the X-ray band, and by Krejcova & Budai (2012) in a survey of Ca II H&K line spectroscopy. On the other hand, Poppenhaeger et al. (2010), by conducting a statistical analysis of all known planet-bearing stars within 30 pc distance, have concluded that coronal SPI might manifest itself only in a few peculiar targets. In HD 189733 Pillitteri et al. (2010, 2011) have claimed that the primary star underwent a boost of activity due to transfer of angular momentum from its hot Jupiter. The same result is found for the binary system of CoRoT-2A by Schroeter et al. (2011). Another evidence for magnetic SPI in HD189733 is the presence of strong flaring activity in a restricted range of planetary phases (Pillitteri et al. 2011) which suggests that the planet could trigger magnetic reconnection along its orbit, as it has been predicted by state of the art MHD modeling of the interaction between star and planet magnetospheres (Cohen et al. 2011). *Athena+* will allow investigating the presence of this flaring activity and other signatures of magnetic SPI in a wider sample of interesting systems currently inaccessible due to their fainter X-ray flux.

Finally, in nearby systems *Athena+* will allow us to search for and study the neutral atom clouds around planets where Charge eXchange (CX), similarly to what has been observed in planets in our Solar System, can take place. The Jupiter-like planet in a very eccentric orbit (P ~ 2500 days) around the nearby (3.2 pc) very active (Log $L_X$ = 28.22) star epsilon Eri, or the 2 (out of 4) planets in eccentric orbits (orbital period in the 2-30 day range) around GL 876 (d = 4.7 pc, Log $L_X$ = 26.48), are promising targets of election for such searches. In GL 876 the planets orbit at a distance of 0.02-0.1 AU, and as a result the incident X-ray flux is from 250 to 6000 times higher that the Sun's coronal flux incident on Jupiter. Even if current predictions based on simple scaling of what we know from solar system studies indicate a very low likelihood of detecting the planet's X-ray emission signatures against the overwhelming stellar X-ray emission, experience with X-ray science over decades suggests that unexpected discoveries are certainly possible (starting from the original detection of stellar X-rays with the Einstein Observatory !).